\documentclass[dvips]{article}
\usepackage{supertabular,lscape,epsfig}
\usepackage{amssymb}
\usepackage{amsmath}
\usepackage{rotating}
\DeclareSymbolFont{ppa}{OT1}{ppl}{m}{it}
\DeclareMathSymbol{\vv}{\mathalpha}{ppa}{'166}

\begin{document}

\newcommand{\dd}{\,{\rm d}}
\newcommand{\ie}{{\it i.e.},\,}
\newcommand{\etal}{{\it et al.\ }}
\newcommand{\eg}{{\it e.g.},\,}
\newcommand{\cf}{{\it cf.\ }}
\newcommand{\vs}{{\it vs.\ }}
\newcommand{\zdot}{\makebox[0pt][l]{.}}
\newcommand{\up}[1]{\ifmmode^{\rm #1}\else$^{\rm #1}$\fi}
\newcommand{\dn}[1]{\ifmmode_{\rm #1}\else$_{\rm #1}$\fi}
\newcommand{\upd}{\up{d}}
\newcommand{\uph}{\up{h}}
\newcommand{\upm}{\up{m}}
\newcommand{\ups}{\up{s}}
\newcommand{\arcd}{\ifmmode^{\circ}\else$^{\circ}$\fi}
\newcommand{\arcm}{\ifmmode{'}\else$'$\fi}
\newcommand{\arcs}{\ifmmode{''}\else$''$\fi}
\newcommand{\MS}{{\rm M}\ifmmode_{\odot}\else$_{\odot}$\fi}
\newcommand{\RS}{{\rm R}\ifmmode_{\odot}\else$_{\odot}$\fi}
\newcommand{\LS}{{\rm L}\ifmmode_{\odot}\else$_{\odot}$\fi}

\newcommand{\Abstract}[2]{{\footnotesize\begin{center}ABSTRACT\end{center}
\vspace{1mm}\par#1\par
\noindent
{~}{\it #2}}}

\newcommand{\TabCap}[2]{\begin{center}\parbox[t]{#1}{\begin{center}
  \small {\spaceskip 2pt plus 1pt minus 1pt T a b l e}
  \refstepcounter{table}\thetable \\[2mm]
  \footnotesize #2 \end{center}}\end{center}}

\newcommand{\TableSep}[2]{\begin{table}[p]\vspace{#1}
\TabCap{#2}\end{table}}

\newcommand{\FigCap}[1]{\footnotesize\par\noindent Fig.\  %
  \refstepcounter{figure}\thefigure. #1\par}

\newcommand{\TableFont}{\footnotesize}
\newcommand{\TableFontIt}{\ttit}
\newcommand{\SetTableFont}[1]{\renewcommand{\TableFont}{#1}}

\newcommand{\MakeTable}[4]{\begin{table}[htb]\TabCap{#2}{#3}
  \begin{center} \TableFont \begin{tabular}{#1} #4
  \end{tabular}\end{center}\end{table}}

\newcommand{\MakeTableSep}[4]{\begin{table}[p]\TabCap{#2}{#3}
  \begin{center} \TableFont \begin{tabular}{#1} #4
  \end{tabular}\end{center}\end{table}}

\newenvironment{references}%
{
\footnotesize \frenchspacing
\renewcommand{\thesection}{}
\renewcommand{\in}{{\rm in }}
\renewcommand{\AA}{Astron.\ Astrophys.}
\newcommand{\AAS}{Astron.~Astrophys.~Suppl.~Ser.}
\newcommand{\ApJ}{Astrophys.\ J.}
\newcommand{\ApJS}{Astrophys.\ J.~Suppl.~Ser.}
\newcommand{\ApJL}{Astrophys.\ J.~Letters}
\newcommand{\AJ}{Astron.\ J.}
\newcommand{\IBVS}{IBVS}
\newcommand{\PASP}{P.A.S.P.}
\newcommand{\PASJ}{PASJ}
\newcommand{\Acta}{Acta Astron.}
\newcommand{\MNRAS}{MNRAS}
\renewcommand{\and}{{\rm and }}
\section{{\rm REFERENCES}}
\sloppy \hyphenpenalty10000
\begin{list}{}{\leftmargin1cm\listparindent-1cm
\itemindent\listparindent\parsep0pt\itemsep0pt}}%
{\end{list}\vspace{2mm}}

\def\TYLDA{~}
\newlength{\DW}
\settowidth{\DW}{0}
\newcommand{\dw}{\hspace{\DW}}

\newcommand{\refitem}[5]{\item[]{#1} #2%
\def\REFARG{#3}\ifx\REFARG\TYLDA\else, {\it#3}\fi
\def\REFARG{#4}\ifx\REFARG\TYLDA\else, {\bf#4}\fi
\def\REFARG{#5}\ifx\REFARG\TYLDA\else, {#5}\fi.}

\newcommand{\Section}[1]{\section{#1}}
\newcommand{\Subsection}[1]{\subsection{#1}}
\newcommand{\Acknow}[1]{\par\vspace{5mm}{\bf Acknowledgments.} #1}
\pagestyle{myheadings}

\newfont{\bb}{ptmbi8t at 12pt}
\newcommand{\xrule}{\rule{0pt}{2.5ex}}
\newcommand{\xxrule}{\rule[-1.8ex]{0pt}{4.5ex}}
\def\thefootnote{\fnsymbol{footnote}}

\begin{center}
{\Large\bf Dwarf Novae in the OGLE Data.\\
II. Forty New Dwarf Novae in the OGLE-III\\ Galactic Disk
Fields\footnote{Based on observations obtained with
the 1.3-m Warsaw telescope at the Las Campanas Observatory
of the Carnegie Institution for Science.}}
\vskip1cm
{\bf
P.~~M~r~\'o~z$^1$,~~P.~~P~i~e~t~r~u~k~o~w~i~c~z$^1$,~~R.~~P~o~l~e~s~k~i$^{1,2}$,\\
~~A.~~U~d~a~l~s~k~i$^1$,~~I.~~S~o~s~z~y~\'n~s~k~i$^1$,~~M.~K.~~S~z~y~m~a~\'n~s~k~i$^1$,\\
~~M.~~K~u~b~i~a~k$^1$,~~G.~~P~i~e~t~r~z~y~\'n~s~k~i$^{1,3}$,~~{\L}.~~W~y~r~z~y~k~o~w~s~k~i$^{1,4}$,\\
~~K.~~U~l~a~c~z~y~k$^1$,~~S.~~K~o~z~{\l}~o~w~s~k~i$^1$ and~~J.~~S~k~o~w~r~o~n$^1$\\}
\vskip3mm
{
$^1$Warsaw University Observatory, Al. Ujazdowskie 4, 00-478 Warsaw, Poland\\
e-mail:
(pmroz,pietruk,rpoleski,udalski,soszynsk,msz,mk,pietrzyn,\\
wyrzykow,kulaczyk,simkoz,jskowron)@astrouw.edu.pl\\
$^2$ Department of Astronomy, Ohio State University, 140 W. 18th Ave.,\\
Columbus, OH 43210, USA\\
$^3$ Universidad de Concepci{\'o}n, Departamento de Fisica,
Casilla 160-C, Concepci{\'o}n, Chile\\
$^4$ Institute of Astronomy, University of Cambridge, Madingley Road,\\
Cambridge CB3 0HA, UK\\
}
\end{center}


\Abstract{We report the discovery of forty erupting cataclysmic variable
stars in the OGLE-III Galactic disk fields: seventeen objects
of U~Gem type, four of Z~Cam type, and nineteen stars showing
outbursts and superoutbursts typical for SU~UMa type dwarf novae.
In the case of five stars we were able to estimate their supercycle lengths.
The obtained lengths are in the range 20--90~d, generally between
the typical SU~UMa type variables and a few objects classified
as the ER~UMa type variables. Since there is no significant
difference between the two types but a higher mass-transfer rate
resulting in more frequent outbursts and superoutbursts in the
ER~UMa type stars, we propose to discard this type as a separate class
of variables. We note that in one of the SU~UMa type stars, OGLE-GD-DN-039,
we found a negative supercycle period change, in contrast to other
active systems of this type. Two of the new OGLE objects showed
long-duration WZ~Sge-like superoutbursts followed by a sequence
of echo outbursts. All stars reported in this paper
are part of the OGLE-III Catalog of Variable Stars.}

{Galaxy: disk -- binaries: close -- novae, cataclysmic variables}


\section{Introduction}

Dwarf novae are eruptive cataclysmic variable (CV) stars. In these close
binary systems, the magnetic field of a white dwarf is weak enough
allowing the matter from a low-mass main-sequence secondary to create
an accretion disk. Under certain conditions, the disk becomes unstable
and the matter falls onto the surface of the white dwarf, releasing
a substantial amount of gravitational energy.

The first dwarf nova (DN), U Gem, was discovered serendipitously
in December 1855 by J.~R. Hind during his search for minor planets
(Warner 1995). By the early 1900's successive discoveries were made,
\eg SS Cyg by L. Wells in 1896, Z~Cam by G. Van Biesbroeck
in 1904, SU~UMa by L. Ceraski in 1908. In 1938 forty-seven
DNe were already known (Payne-Gaposchkin \& Gaposchkin 1938).
The invention of photoelectric photometry in 1940s and application
of CCD detectors in 1990s led to further discoveries and in-depth
studies of CVs. G\"{a}nsicke (2005) lists the most important photometric
surveys, including the Palomar-Green survey (Green \etal 1986),
the Edinburgh-Cape survey (Stobie \etal 1997), the Hamburg Quasar Survey
(Hagen \etal 1995), and the 2dF Quasar Survey (Boyle \etal 2000), which
increased the number of known CVs to several hundreds.

Recently, a large number of cataclysmic variables has been discovered mainly
in the course of two projects. Based on the data from Sloan Digital Sky Survey
(SDSS, York \etal 2000) 285 CVs were identified (Szkody \etal 2011), most
of which had not been known before. These objects were selected using their
colors and confirmed spectroscopically. Additional observations showed
the presence of at least nine DN candidates in that sample
(Szkody \etal 2011). In the second project, the Catalina Real-time Transient
Survey (CRTS), numerous optical transient events, including DN outbursts,
supernovae, blazars, UV~Cet type flares, are being detected
(Drake \etal 2012). So far, by June 2013, the CRTS
website\footnote{http://nesssi.cacr.caltech.edu/catalina/AllCV.html}
lists around 690 confirmed and candidate CVs. From time to time new CVs
and candidates for this type of variables are being detected by other optical
wide-field surveys such as the Palomar Transient Factory (PTF, Law \etal 2009),
the All Sky Automated Survey (Pojma\'nski 1997), the OGLE Transient Survey
(Koz{\l}owski \etal 2013).

However, none of the mentioned above surveys concentrate on crowded stellar
regions such as the Galactic disk and bulge. In the bulge area, Cieslinski
\etal (2003) and Cieslinski \etal (2004) found 33 and 28 DNe,
respectively, searching the OGLE-II and MACHO databases for brightenings
in the light curves. Recently, Poleski \etal (2011) reported the
identification of three new DNe detected in the OGLE-IV bulge area and
presented potential of large-field high-cadence surveys for DN studies.
In Fig.~1, we show the distribution of known DNe in the sky.

Based on the observed light variations DNe are divided into three main
types: U~Gem, SU~UMa, and Z~Cam variables.

U Gem type stars (also called SS Cyg type stars) exhibit
quasi-regular (with recurrent times from several days to several years)
large-amplitude (2--5~mag in $V$) outbursts. According to the standard
model (Osaki 1974) the thermal instability in the accretion disk
triggers the matter to fall onto the surface of the white dwarf.
In this process a substantial amount of gravitational energy is released
and the luminosity of the system increases by up to $\approx100$ times.
The orbital periods of U~Gem type variables are usually longer than 3~h.

SU~UMa type stars have occasional superoutbursts brighter
and longer than normal outbursts. During the superoutbursts light
variations of an amplitude up to $\approx0.4$~mag in the $V$ band,
called superhumps, are observed. Currently, there are three competitive
models explaining the behavior of SU~UMa type variables: (1) the thermal-tidal
instability (TTI) model (Osaki 1989), in which the ordinary thermal
instability is coupled with the tidal instability and 
superhumps are the result of an eccentric disk; (2) the enhanced
mass-transfer (EMT) model (Smak 1991, 2004, 2008), in which the mass-transfer
rate from the secondary is variable and superhumps are due to variable
brightness of the hot spot at the edge of the disk; and (3) the pure
thermal limit cycle model (Cannizzo \etal 2010, 2012), in which the thermal
instability is complex enough to produce superoutbursts and supercycles.
According to the very recent work by Osaki and Kato (2013) the short cadence
data from the orbiting Kepler telescope (Koch \etal 2010)
of DNe V344 Lyr and V1504 Cyg strongly favors the TTI model.
Typical SU~UMa type DNe have supercycle lengths $P_{\rm sc}$
between $\approx100$~d and $\approx1000$~d. Dwarf novae with
$P_{\rm sc}\lesssim60$~d (so far five objects) have been classified
as ER~UMa type variables (Otulakowska-Hypka \etal 2013a), while stars
with $P_{\rm sc}\gtrsim3000$~d are known as WZ~Sge type variables
(Uemura \etal 2010). The orbital periods of SU~UMa type stars are shorter
than 3~h.

Dwarf novae of the Z~Cam type exhibit standstills,
during which outbursts cease for days to years. In standstill
the star is up to one magnitude below its outburst level.
Meyer and Meyer-Hofmeister (1983) proposed that Z~Cam type stars
have mass-transfer rates just below the critical value
in cycling outbursting phases and slightly above in standstills,
as it is in the case of nova-like stars.

In this paper we report the discovery of forty DNe in the OGLE-III
Galactic disk fields. Section~2 gives details on the observations
and reductions. The search procedures and the analysis of light curves are
described in Section~3 and Section~4, respectively. In the subsequent
three sections we present the results on stars classified to the
U~Gem type (Section~5), SU~UMa type (Section~6), and Z~Cam type
(Section~7). Finally, Section~8 states our conclusions.

\begin{figure}[htb]
\centerline{\includegraphics[angle=270,width=130mm]{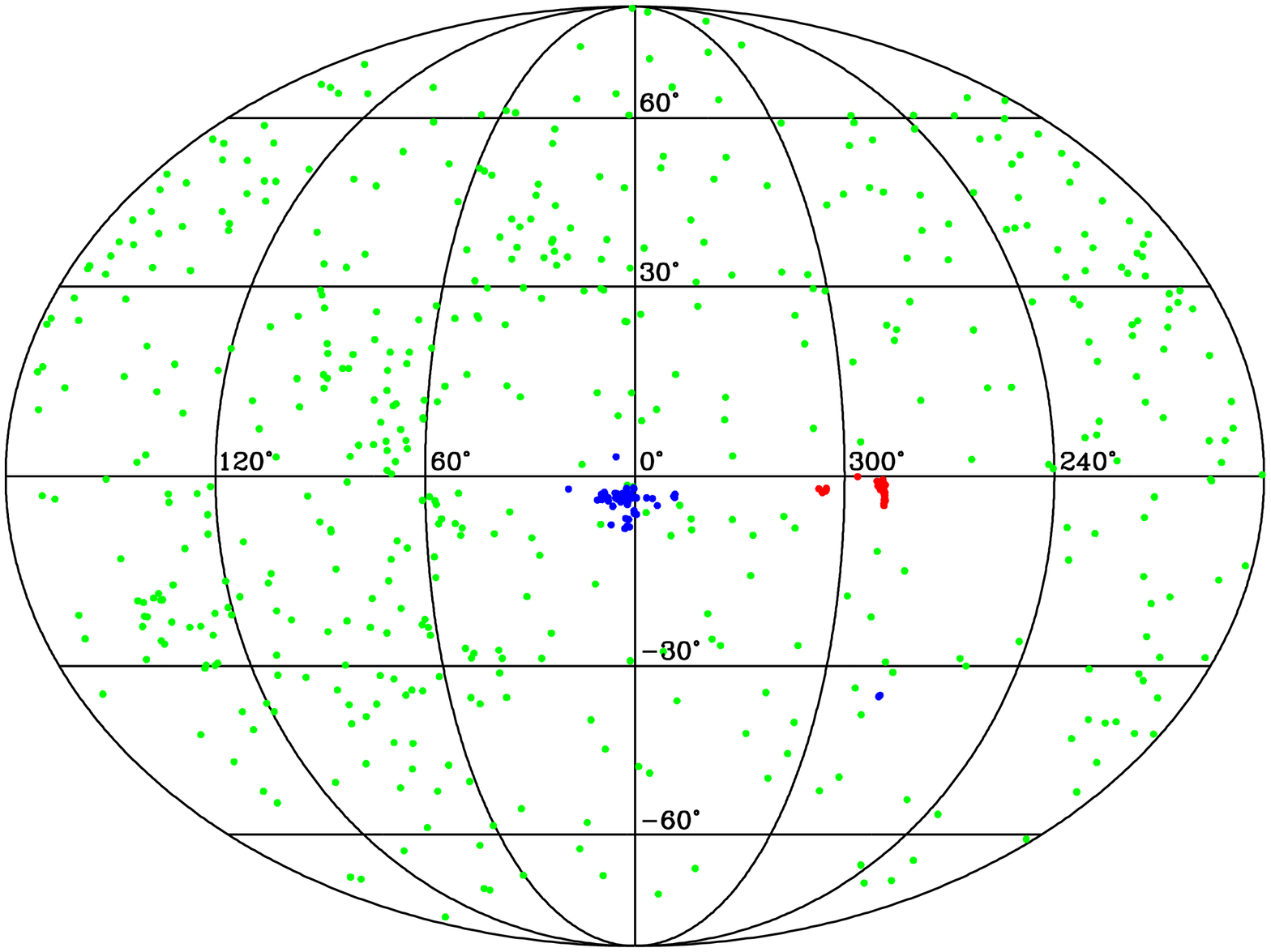}}
\FigCap{Distribution, in the Galactic coordinates, of known dwarf novae.
Green points denote positions of 507 DNe listed in the 2013 version
of the Ritter \& Kolb catalog (Ritter \& Kolb 2003), blue points
represent 65 DNe found in the OGLE-II and MACHO Galactic bulge data
(Cieslinski \etal 2003 and Cieslinski \etal 2004, respectively), while
red points show positions of 40 new DNe reported in this paper.}
\end{figure}


\section{Observations and Data Reductions}

All data presented in this paper were collected between 2001 and 2009
during the third phase of the Optical Gravitational Lensing Experiment
(OGLE-III). The observations were obtained with the 1.3-m Warsaw Telescope
located at Las Campanas Observatory, Chile, which is operated by Carnegie
Institution for Science. A CCD camera attached to the telescope consisted
of eight chips with the total field of view of $35\arcm \times 35\arcm$
and the scale of 0.26 arcsec/pixel. More details concerning the
instrumentation can be found in Udalski (2003).

The OGLE-III Galactic disk fields cover an area of 7.12~deg$^2$
toward the constellations of Carina, Centaurus, and Musca. The time-span
of the observations and the number of data points varied from
field to field (see Table~1 and Figs.~1--2 in Pietrukowicz \etal 2013).
A total number of approximately $8.8 \cdot 10^6$ stars was observed
(Szyma\'nski \etal 2010) in these lines-of-site.

A majority of observations were taken in the standard $I$-band filter with
an exposure time of 120 or 180~s. For the remaining measurements,
the $V$-band filter was used with a longer exposure time of 240~s.
The photometry was carried out with the Difference Image Analysis (DIA)
algorithm (Alard and Lupton 1998, Wo\'zniak 2000). A detailed description
of the OGLE data reduction pipeline can be found in Udalski \etal (2008).

In the case of eighteen faint DNe, we decided to improve their photometry.
For this purpose we selected subtracted images taken during outbursts
of the DNe under good seeing conditions. After stacking the images,
we re-determined positions of the centroids for these stars. Finally, having
the new centroids, we extracted the photometry using the standard pipeline.

The final $I$-band light curves of the newly detected dwarf novae are
available to the astronomical community from the OGLE Internet Archive:
\begin{center}
{\it http://ogle.astrouw.edu.pl\\
ftp://ftp.astrouw.edu.pl/ogle/ogle3/OIII-CVS/gd/dn/\\}
\end{center}
The objects are arranged according to increasing right ascension
and named as OGLE-GD-DN-NNN, where NNN is a consecutive number.
We note that DNe reported in this paper are included in
the OGLE-III Catalog of Variable Stars.


\section{Search for Dwarf Novae}

Dwarf novae exhibit a vast diversity of photometric behavior. Their light
curve shapes often differ much even for members of the same subgroup.
Outbursts appear unexpectedly, and in many cases, they recur
non-periodically. All these factors make the search for dwarf novae
very demanding. Our search procedure consisted of three independent parts.

In the beginning, 27 DNe were discovered as a byproduct of the search
for periodic variable stars in the OGLE-III Galactic disk fields
(Pietrukowicz \etal 2013). Phased and time-domain light curves of almost
345 500 stars with a signal to noise S/N$>10$ were visually inspected
resulting in thousands of periodic variables such as eclipsing
binaries (Pietrukowicz \etal 2013), pulsating stars, and spotted
stars (Pietrukowicz \etal in preparation). We identified the first 27 DNe,
mainly with large amplitudes and quasi-periodic brightenings, among
miscellaneous variables. This relatively large number of DNe encouraged
us to further investigations.

It is commonly known that the majority of cataclysmic variables
and particularly dwarf novae exhibit UV and blue excess (Warner 1995).
Unfortunately, the observations were collected only in the optical $V$-
and $I$-bands. Due to large interstellar absorption in the observed
disk fields, this makes a color-based selection of CVs/DNe practically
impossible. On the other hand, we decided to examine the optical
counterparts to the X-ray sources detected within the OGLE-III
disk area (Balman 2012, Udalski \etal 2012). According to the HEASARC
(the High Energy Astrophysics Science Archive Research
Center\footnote{http://heasarc.gsfc.nasa.gov/}) database there are
220 such sources. We searched for optically variable candidates
within the radius of $5\zdot\arcs2$ ($=20$~pix) around their positions.
Only one dwarf nova, OGLE-GD-DN-030, was found $0\zdot\arcs47$ away
from the source X110505.31-611044.5 (Ptak \& Griffiths 2003).
Incidentally, we note that this source is located $3\zdot\arcm8$ away
from the pulsar PSR J1105-6107, which was observed with the Chandra
X-Ray Observatory twice in 2002. This result shows that the current
X-ray surveys are shallow and/or limited to small regions in the sky.

We were aware that some DNe might have been overlooked during
the visual inspection or they simply did not pass the periodicity
condition (\eg only one outburst was observed). In order to increase
the completeness as much as possible, we applied the following procedure.
For all $\approx8.8 \cdot 10^6$ stars with the available $I$-band
photometry observations were grouped into one-day bins and the
mean brightness per night with at least three data points was calculated.
In the first approach, DN candidates were selected if their
brightness increased by at least 1.0~mag over the mean value for at least
three consecutive days. From about 12 500 visually inspected
candidates we found twelve new dwarf novae. In the second approach,
we extended the search for variables with amplitudes at least 0.8~mag
and a mean brightness $I<19.5$~mag, but no new objects were found.
Our algorithm found 24 DNe already detected in the first part. It did not
recognize two Z~Cam type stars and two SU~UMa type variables with amplitudes
between 0.3 and 0.8~mag, which are very likely blended objects.
Apart from the {\it bona fide} DNe, the majority of candidates were 
DY~Per type stars, Be stars, long period variables, and artifacts.

The three independent searches resulted in the discovery of
40 previously unknown DNe. None of them is listed
neither in the Ritter \& Kolb catalog (Ritter \& Kolb 2003,
RKcat Edition 7.19, 2013) nor in the General Catalogue of Variable Stars,
(GCVS, Samus \etal, 2007--2012). In Table~1, we give basic
information on the discovered DNe. Figs.~2--3 present finding
charts for our objects.

\begin{table}[h!]
\centering
\caption{\small Basic data on the discovered DNe in the OGLE-III disk fields}
\medskip
{\small
\begin{tabular}{ccrccl}
\hline
Name & Field & Star ID & RA & Dec & Type\\
     &       &         & J2000.0 & J2000.0 & \\
\hline
OGLE-GD-DN-001 & CAR118.8 &  1241 & $10\uph36\upm19\zdot\ups78$ & $-63\arcd37\arcm32\zdot\arcs8$ & SU/WZ \\
OGLE-GD-DN-002 & CAR116.2 & 10591 & $10\uph37\upm28\zdot\ups40$ & $-62\arcd49\arcm28\zdot\arcs8$ & UG \\
OGLE-GD-DN-003 & CAR118.7 & 29174 & $10\uph38\upm00\zdot\ups57$ & $-63\arcd22\arcm47\zdot\arcs7$ & SU \\
OGLE-GD-DN-004 & CAR116.3 & 27971 & $10\uph38\upm45\zdot\ups69$ & $-62\arcd38\arcm52\zdot\arcs0$ & SU \\
OGLE-GD-DN-005 & CAR116.3 &  6361 & $10\uph38\upm56\zdot\ups17$ & $-62\arcd43\arcm15\zdot\arcs8$ & UG \\
OGLE-GD-DN-006 & CAR117.8 & 28538 & $10\uph39\upm47\zdot\ups75$ & $-62\arcd55\arcm29\zdot\arcs4$ & UG \\
OGLE-GD-DN-007 & CAR117.7 & 31482 & $10\uph39\upm57\zdot\ups00$ & $-62\arcd45\arcm17\zdot\arcs8$ & SU \\
OGLE-GD-DN-008 & CAR117.5 & 13030 & $10\uph40\upm14\zdot\ups25$ & $-62\arcd33\arcm17\zdot\arcs7$ & SU \\
OGLE-GD-DN-009 & CAR109.7 & 35705 & $10\uph42\upm02\zdot\ups40$ & $-62\arcd11\arcm53\zdot\arcs5$ & SU \\
OGLE-GD-DN-010 & CAR110.1 & 28587 & $10\uph43\upm30\zdot\ups62$ & $-61\arcd47\arcm36\zdot\arcs8$ & UG \\
OGLE-GD-DN-011 & CAR109.3 & 16569 & $10\uph43\upm34\zdot\ups50$ & $-62\arcd07\arcm34\zdot\arcs4$ & UG \\
OGLE-GD-DN-012 & CAR107.5 &  5148 & $10\uph46\upm16\zdot\ups97$ & $-61\arcd50\arcm56\zdot\arcs8$ & ZC \\
OGLE-GD-DN-013 & CAR111.3 & 38101 & $10\uph47\upm46\zdot\ups92$ & $-60\arcd41\arcm08\zdot\arcs8$ & UG \\
OGLE-GD-DN-014 & CAR111.3 & 39243 & $10\uph47\upm57\zdot\ups16$ & $-60\arcd41\arcm22\zdot\arcs7$ & SU/WZ \\
OGLE-GD-DN-015 & CAR108.1 & 17123 & $10\uph47\upm58\zdot\ups43$ & $-61\arcd38\arcm17\zdot\arcs7$ & SU \\
OGLE-GD-DN-016 & CAR108.2 & 16574 & $10\uph48\upm07\zdot\ups40$ & $-61\arcd30\arcm13\zdot\arcs5$ & SU \\
OGLE-GD-DN-017 & CAR105.8 & 42407 & $10\uph50\upm16\zdot\ups68$ & $-61\arcd50\arcm09\zdot\arcs1$ & SU \\
OGLE-GD-DN-018 & CAR105.8 & 44632 & $10\uph50\upm47\zdot\ups10$ & $-61\arcd50\arcm09\zdot\arcs7$ & UG \\
OGLE-GD-DN-019 & CAR105.5 & 24506 & $10\uph51\upm36\zdot\ups65$ & $-61\arcd27\arcm39\zdot\arcs1$ & UG \\
OGLE-GD-DN-020 & CAR105.7 & 51393 & $10\uph52\upm11\zdot\ups85$ & $-61\arcd41\arcm14\zdot\arcs3$ & UG \\
OGLE-GD-DN-021 & CAR105.3 & 33886 & $10\uph52\upm49\zdot\ups48$ & $-61\arcd34\arcm16\zdot\arcs5$ & UG \\
OGLE-GD-DN-022 & CAR105.4 & 38638 & $10\uph54\upm00\zdot\ups44$ & $-61\arcd24\arcm38\zdot\arcs3$ & UG \\
OGLE-GD-DN-023 & CAR104.7 &  1357 & $10\uph55\upm03\zdot\ups97$ & $-61\arcd56\arcm52\zdot\arcs8$ & SU \\
OGLE-GD-DN-024 & CAR114.5 &  2145 & $10\uph55\upm20\zdot\ups52$ & $-60\arcd17\arcm28\zdot\arcs1$ & UG \\
OGLE-GD-DN-025 & CAR113.6 & 11221 & $10\uph57\upm17\zdot\ups30$ & $-61\arcd02\arcm04\zdot\arcs4$ & SU \\
OGLE-GD-DN-026 & CAR104.7 & 60634 & $10\uph57\upm30\zdot\ups00$ & $-61\arcd40\arcm30\zdot\arcs3$ & UG \\
OGLE-GD-DN-027 & CAR114.1 & 36099 & $10\uph57\upm50\zdot\ups90$ & $-60\arcd37\arcm22\zdot\arcs6$ & ZC \\
OGLE-GD-DN-028 & CAR113.1 & 31667 & $10\uph59\upm25\zdot\ups08$ & $-61\arcd17\arcm27\zdot\arcs1$ & ZC \\
OGLE-GD-DN-029 & CAR106.6 & 38520 & $11\uph02\upm35\zdot\ups82$ & $-61\arcd44\arcm09\zdot\arcs4$ & UG \\
OGLE-GD-DN-030 & CAR100.7 & 27497 & $11\uph05\upm05\zdot\ups36$ & $-61\arcd10\arcm44\zdot\arcs8$ & UG \\
OGLE-GD-DN-031 & CAR100.2 & 37818 & $11\uph07\upm20\zdot\ups59$ & $-61\arcd08\arcm32\zdot\arcs9$ & SU \\
OGLE-GD-DN-032 & CEN107.8 & 52448 & $11\uph53\upm56\zdot\ups20$ & $-62\arcd12\arcm10\zdot\arcs6$ & UG \\
OGLE-GD-DN-033 & MUS100.5 &  4312 & $13\uph12\upm50\zdot\ups74$ & $-64\arcd39\arcm51\zdot\arcs2$ & SU \\
OGLE-GD-DN-034 & MUS100.6 &  5148 & $13\uph12\upm53\zdot\ups85$ & $-64\arcd49\arcm55\zdot\arcs6$ & SU \\
OGLE-GD-DN-035 & MUS100.1 & 65558 & $13\uph15\upm08\zdot\ups56$ & $-65\arcd00\arcm23\zdot\arcs8$ & ZC \\
OGLE-GD-DN-036 & MUS100.1 & 25683 & $13\uph15\upm29\zdot\ups03$ & $-65\arcd05\arcm38\zdot\arcs0$ & SU \\
OGLE-GD-DN-037 & MUS101.8 & 64830 & $13\uph22\upm16\zdot\ups09$ & $-65\arcd08\arcm00\zdot\arcs5$ & SU \\
OGLE-GD-DN-038 & MUS101.7 & 38066 & $13\uph23\upm57\zdot\ups46$ & $-65\arcd04\arcm35\zdot\arcs7$ & SU \\
OGLE-GD-DN-039 & MUS101.8 & 33967 & $13\uph24\upm01\zdot\ups62$ & $-65\arcd12\arcm19\zdot\arcs8$ & SU \\
OGLE-GD-DN-040 & CEN108.8 & 42697 & $13\uph32\upm39\zdot\ups50$ & $-64\arcd28\arcm35\zdot\arcs9$ & UG \\
\hline
\noalign{\vskip3pt}
\multicolumn{6}{p{12cm}}{\footnotesize Object OGLE-GD-DN-009 also refers
to star 28443 in field CAR115.2.}
\end{tabular}}
\end{table}

\begin{figure}[h!]
\centerline{\includegraphics[angle=0,width=\textwidth]{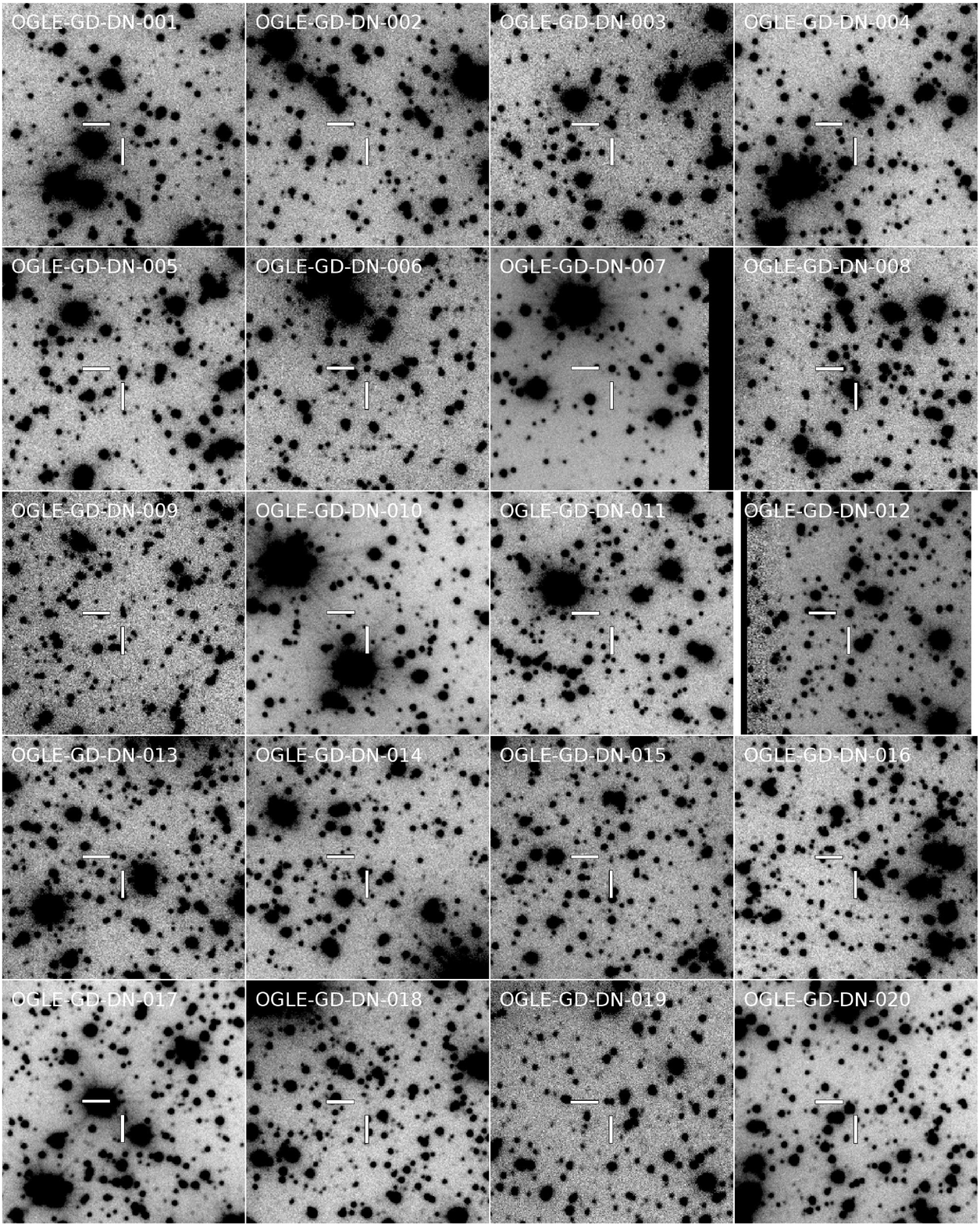}}
\FigCap{Finding charts for the newly discovered DNe in the OGLE-III
disk data (part 1 of 2). North is to the right and East is up.
The field of view is $1\arcm$ on a side. The variables are exactly
in the centers of the charts and marked with cross-hairs.}
\label{fig:spa}
\end{figure}

\begin{figure}[h!]
\centerline{\includegraphics[angle=0,width=\textwidth]{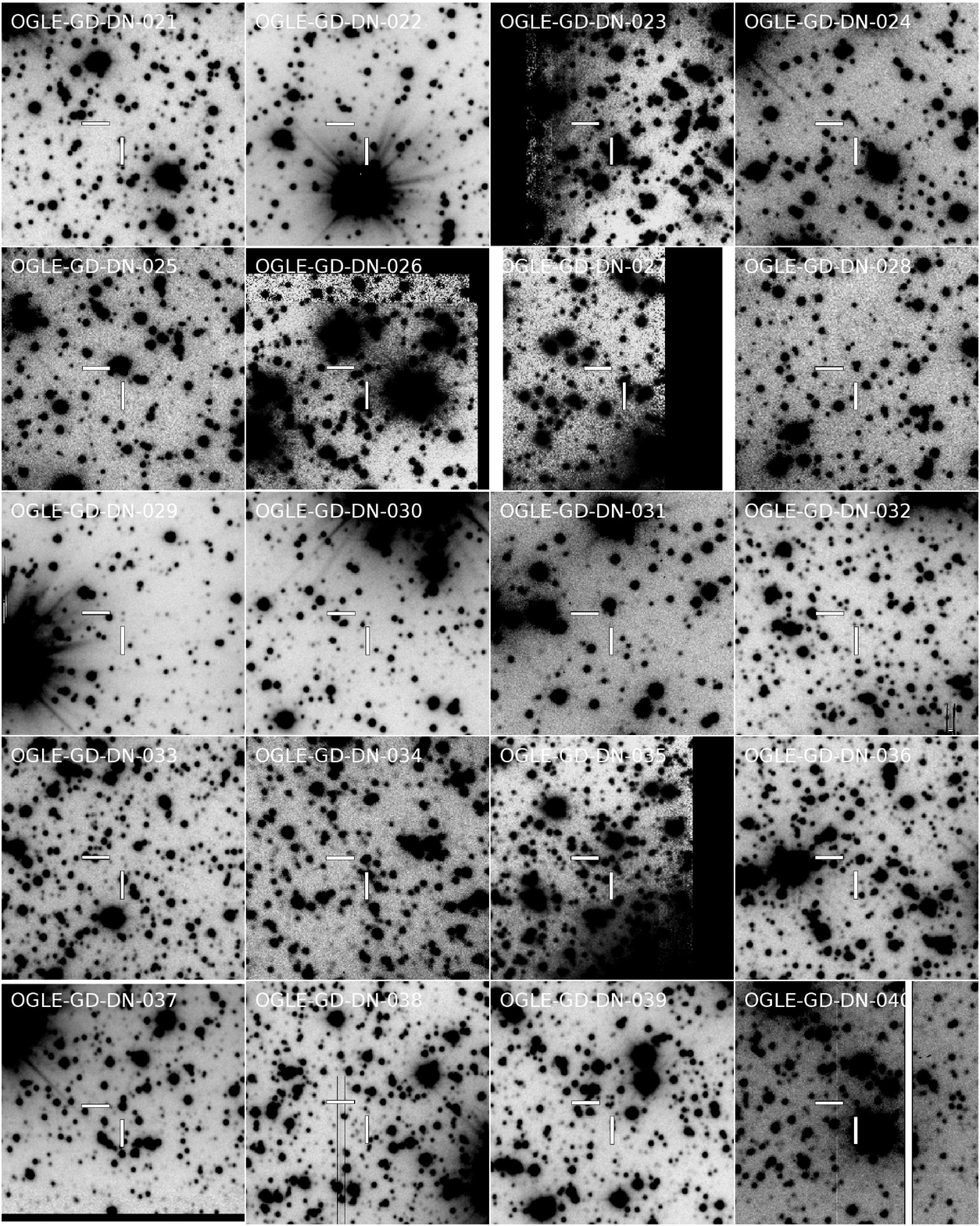}}
\FigCap{Finding charts for the newly discovered DNe in the OGLE-III
disk data (part 2 of 2). North is to the right and East is up.
The field of view is $1\arcm$ on a side. The variables are exactly
in the centers of the charts and marked with cross-hairs.}
\label{fig:spa}
\end{figure}


\section{Light Curve Analysis}

We cleaned the raw light curves of our DNe removing a few obvious outliers
and all data points fainter than $I=23$~mag. Observations were calibrated
on the basis of the mean magnitudes given in the OGLE-III photometric maps
of the Galactic disk (Szyma\'nski \etal 2010). Each final light curve
contains from 678 (OGLE-GD-DN-001) to 3789 (OGLE-GD-DN-009) data
points with the mean value of 1752 points. The time span of the
observations varies from 1 to 7 years, depending on the field.

Each light curve was divided into parts corresponding to quiescence,
outbursts, superoutbursts, and standstills, if those were present.
The magnitudes were transformed into flux and the weighted mean was
calculated for quiescence and standstills. We note that the statistical error
is small (typically $\sigma_I\lesssim0.05$~mag) due to the large number
of measurements. In a similar way, the peak brightness was assessed for
normal outbursts and superoutbursts of SU~UMa type variables. We measured
the time duration of the outbursts and calculated their average value.
In many cases the beginning or/and the end of outburst was not observed.
For some SU~UMa type stars we were able to find the recurrence times
of normal outbursts ($T_{\rm n}$) and superoutbursts ($T_{\rm s}$).
In a few cases the superoutbursts were well-sampled and bright
enough that it was possible to determine the superhump period
after linear de-trending. For three U~Gem type stars we were able
to estimate their orbital periods in quiescence. This was performed with
the help of the {\sc TATRY} code (Schwarzenberg-Czerny 1996).


\section{U Gem type Stars}

\begin{table}
\centering
\caption{\small Photometric data on the detected U~Gem type stars}
\medskip
\begin{tabular}{cccrrrc}
\hline
Name & $I_{\rm qui}$ & $I_{\rm out}$ & $n_{\rm out}$ & $D_{\rm out}$ & $T_{\rm out}$ & $P_{\rm orb}$\\
 & [mag] & [mag] & & [d] & [d] & [d] \\
\hline
OGLE-GD-DN-002 & 20.18 & 18.46 & 22 & 10.0 & 10.7 & 0.177939(3) \\
OGLE-GD-DN-005 & 19.56 & 17.71 & 22 &  6.6 & 12.6 & - \\
OGLE-GD-DN-006 & 19.23 & 17.74 & 33 &  3.0 &  8.6 & 0.269963(9) \\
OGLE-GD-DN-010 & 20.99 & 19.03 &  7 &  9.0 & 11.5 & - \\
OGLE-GD-DN-011 & 20.70 & 18.95 &  3 & 11.0 &    - & - \\
OGLE-GD-DN-013 & 20.51 & 19.13 &  3 & 21.0 & 81.0 & - \\
OGLE-GD-DN-018 & 21.21 & 18.95 &  3 &  5.0 &    - & - \\
OGLE-GD-DN-019 & 21.02 & 18.63 &  7 & 12.1 & 29.0 & 0.426916(3) \\
OGLE-GD-DN-020 & 20.87 & 18.95 &  5 & 13.0 & 54.5 & 0.293191(3) \\
OGLE-GD-DN-021 & 22.19 & 19.78 &  1 &  4.0 &    - & - \\
OGLE-GD-DN-022 & 21.23 & 18.45 &  1 & 15.5 &    - & - \\
OGLE-GD-DN-024 & 21.12 & 19.34 &  4 & 13.0 & 61.0 & - \\
OGLE-GD-DN-026 & 19.38 & 17.34 &  3 & 10.0 & 52.8 & - \\
OGLE-GD-DN-029 & 21.19 & 19.47 &  3 &  8.0 & 27.2 & - \\
OGLE-GD-DN-030 & 19.99 & 17.38 &  1 & $>8.0$ &  - & - \\
OGLE-GD-DN-032 & 20.35 & 18.16 &  1 &  8.0 &    - & - \\
OGLE-GD-DN-040 & 20.70 & 18.63 &  1 &  6.0 &    - & - \\
\hline
\end{tabular}
\end{table}

In Table~2 we give basic photometric information on our U~Gem type
stars: the mean $I$-band brightness in quiescence $I_{\rm qui}$,
the mean $I$-band maximum brightness in outbursts $I_{\rm out}$,
the number of observed outbursts, the duration of outbursts $D_{\rm out}$,
the recurrence time between the outbursts $T_{\rm out}$,
and the orbital period $P_{\rm orb}$. Figs.~4--5 present the light curves
of DNe classified as U~Gem type variables. Phased light curves for three
systems with orbital periods estimated in quiescence are shown in Fig.~6.

\begin{figure}
\centerline{\includegraphics[angle=0,width=100mm]{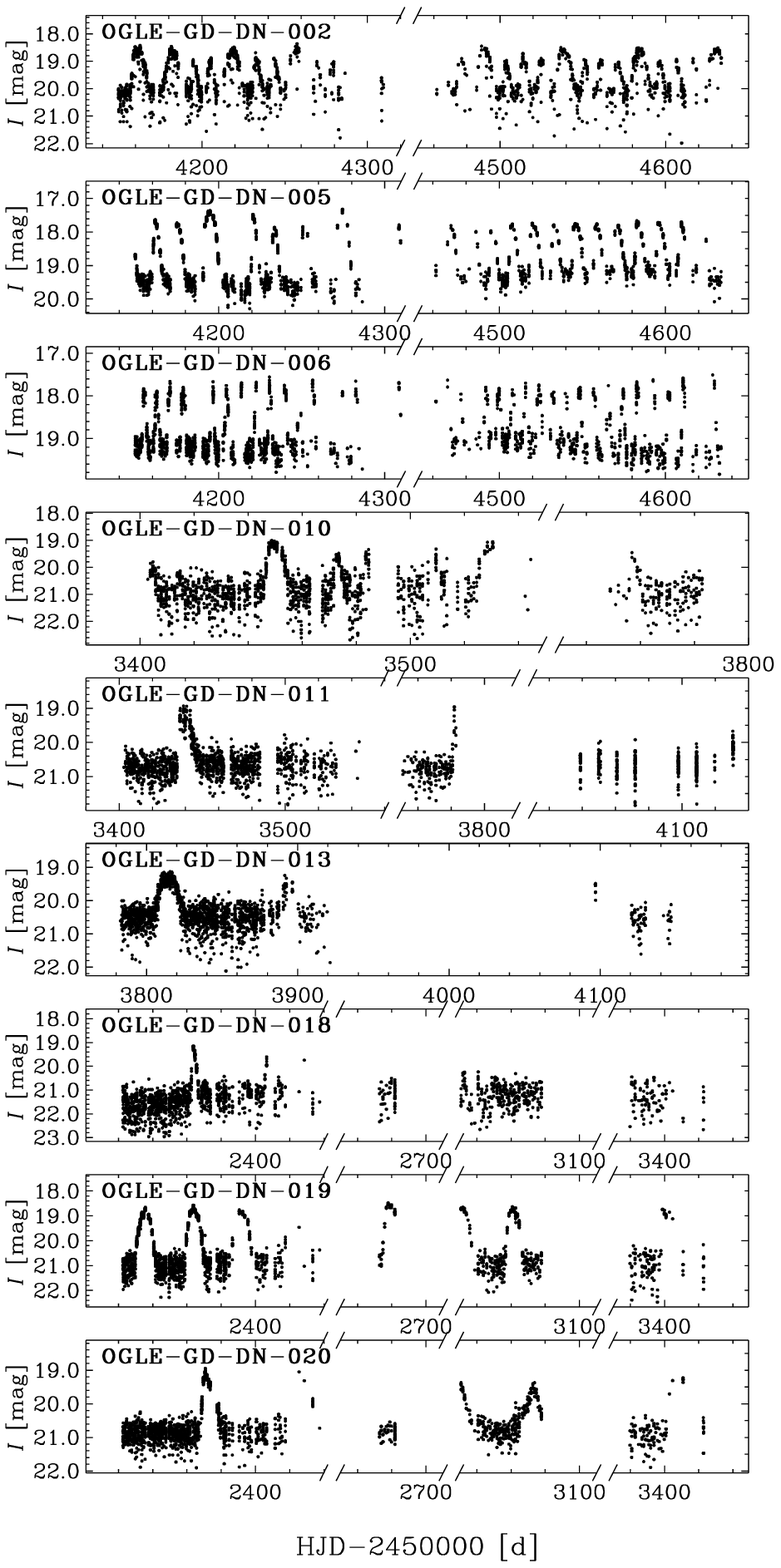}}
\FigCap{Light curves of the U Gem type stars found in the OGLE-III
disk area (part 1 of 2). Small ticks on the time axis are every 20~d.}
\end{figure}

\begin{figure}
\centerline{\includegraphics[angle=0,width=100mm]{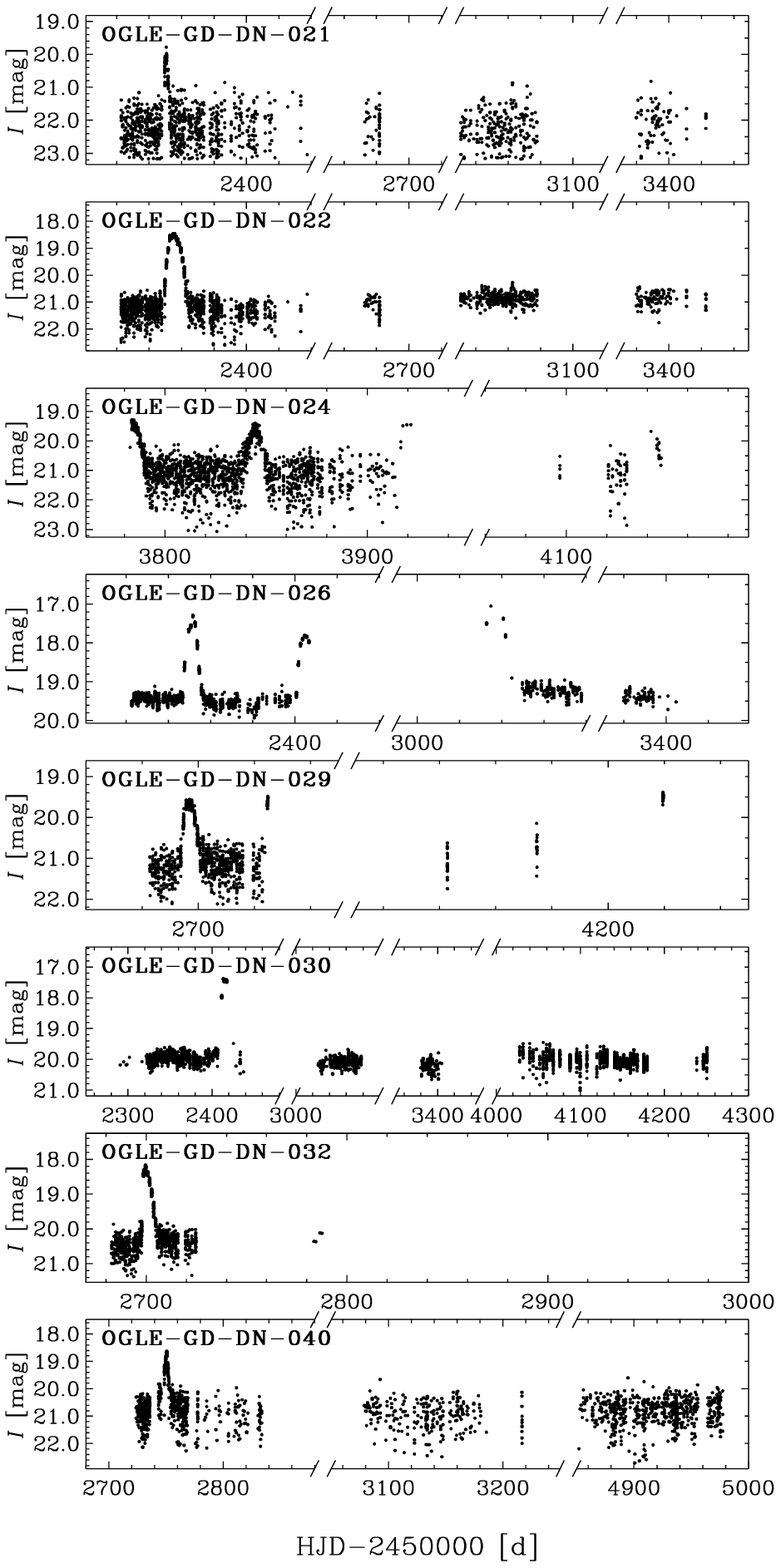}}
\FigCap{Light curves of the U Gem type stars found in the OGLE-III
disk area (part 2 of 2). Small ticks on the time axis are every 20~d.
Objects with the measured brightness of $I\approx22$--$23$~mag
in quiescence (\eg OGLE-GD-DN-021) could be fainter in baseline.}

\end{figure}

\begin{figure}
\centerline{\includegraphics[angle=0,width=100mm]{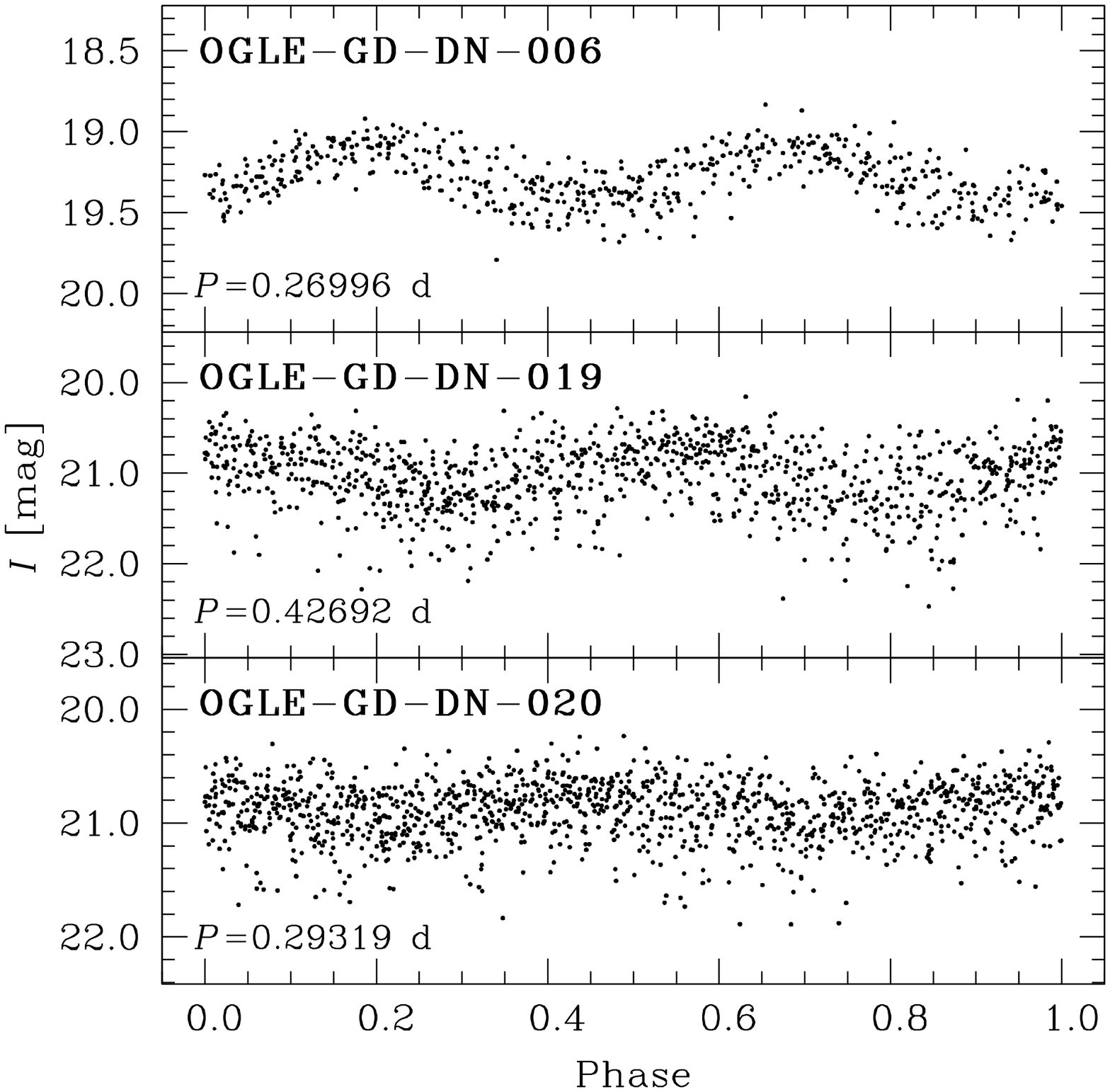}}
\FigCap{Phased light curves of three U Gem type stars in quiescence.}
\end{figure}

\subsection{OGLE-GD-DN-002}

This star was observed during two seasons: 2007 and 2008.
Its global light curve is shown in the most
upper panel of Fig.~4. The scattered data points under the main
light curve are real variations due to eclipses. In order to find
the orbital period, we performed an $O-C$ analysis for two seasons
separately. From the light curve we selected the data points corresponding
to the primary minimum. We adopted a trial period of 0.178~d and calculated
a linear ephemeris. The best period values are 0.177937(3)~d
and 0.177942(4)~d for the 2007 and 2008 season, respectively.
The estimated values do not differ statistically, therefore the derived
orbital period is $P_{\rm orb}=0.177939(3)$~d $=4.27053(6)$~h
and the upper limit for the period derivative is
$\dot{P}_{\rm orb}=+2\cdot 10^{-8}$. The obtained orbital period
places OGLE-GD-DN-002 well above the period gap, in contrast to what
one would expect from its global light curve at the first glance.
With the bright (long) and faint (short) outbursts it might resemble
an SU-UMa type variable, but it is not.

We detrended and phased the light curve during the short outbursts,
long outbursts, and in quiescence (see Fig.~7). During outbursts
the primary eclipses are deeper ($\approx1.8$~mag) than in quiescence
($\approx1.0$~mag), which can be easily explained: during outbursts the
temperature of the accretion disk rises. Simultaneously, in that phase,
the depth of the secondary eclipse decreases. During bright and long
outbursts, the secondary eclipses are practically not visible.

We also report the presence of an embedded precursor in one of the
long outbursts in OGLE-GD-DN-002 (see Fig.~8). Embedded precursors
have been found in V447 Lyr based on the Kepler data (Ramsay \etal 2012)
and in archival data of U~Gem and SS~Cyg (Cannizzo 2012).
Studies of such features may help in understanding of the mechanism
of long outbursts in DNe in general.

\begin{figure}[htb!]
\centerline{\includegraphics[angle=0,width=130mm]{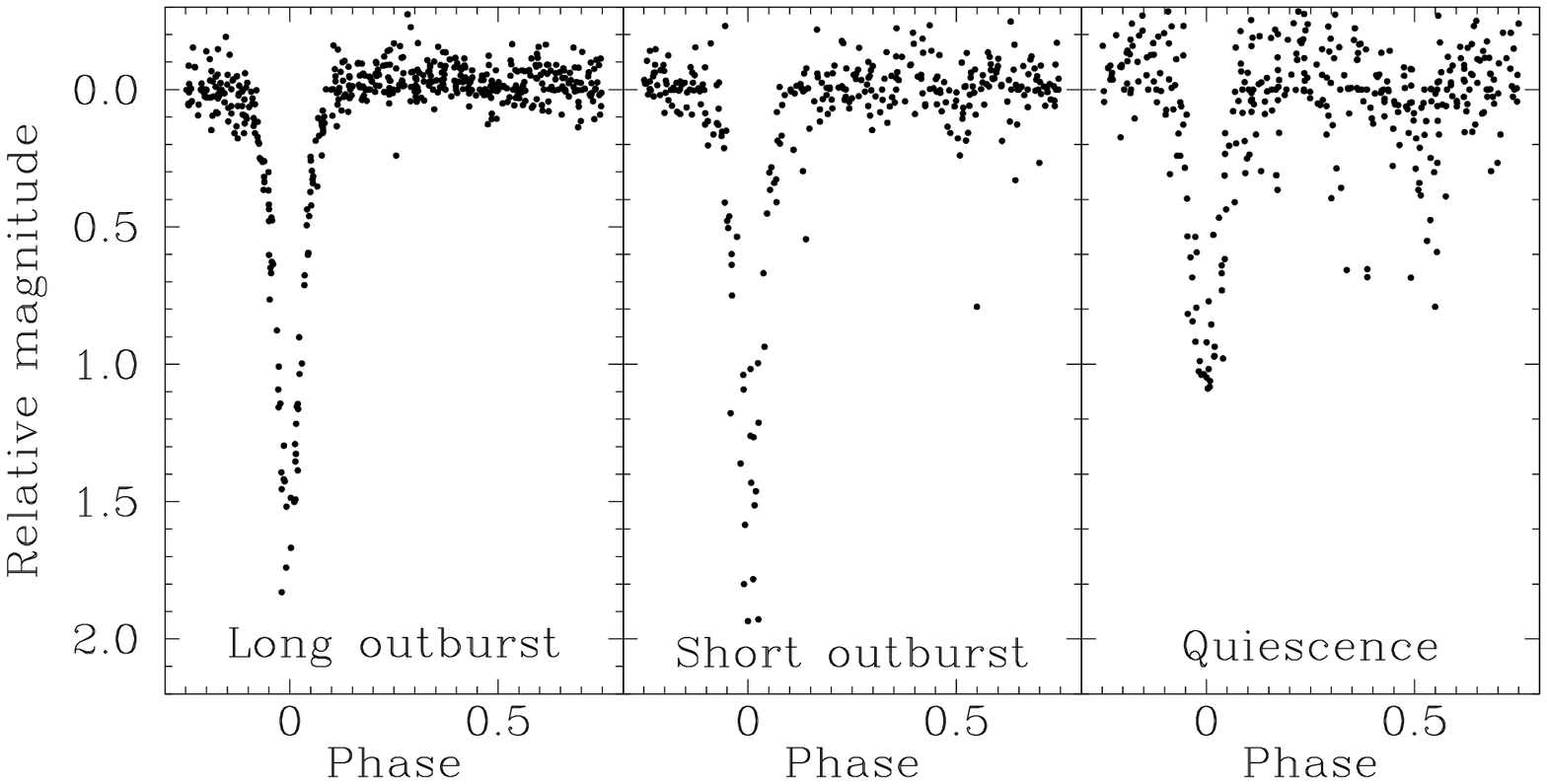}}
\FigCap{Eclipses in OGLE-GD-DN-002 in three different states.}
\end{figure}

\begin{figure}[h!]
\centerline{\includegraphics[angle=0,width=110mm]{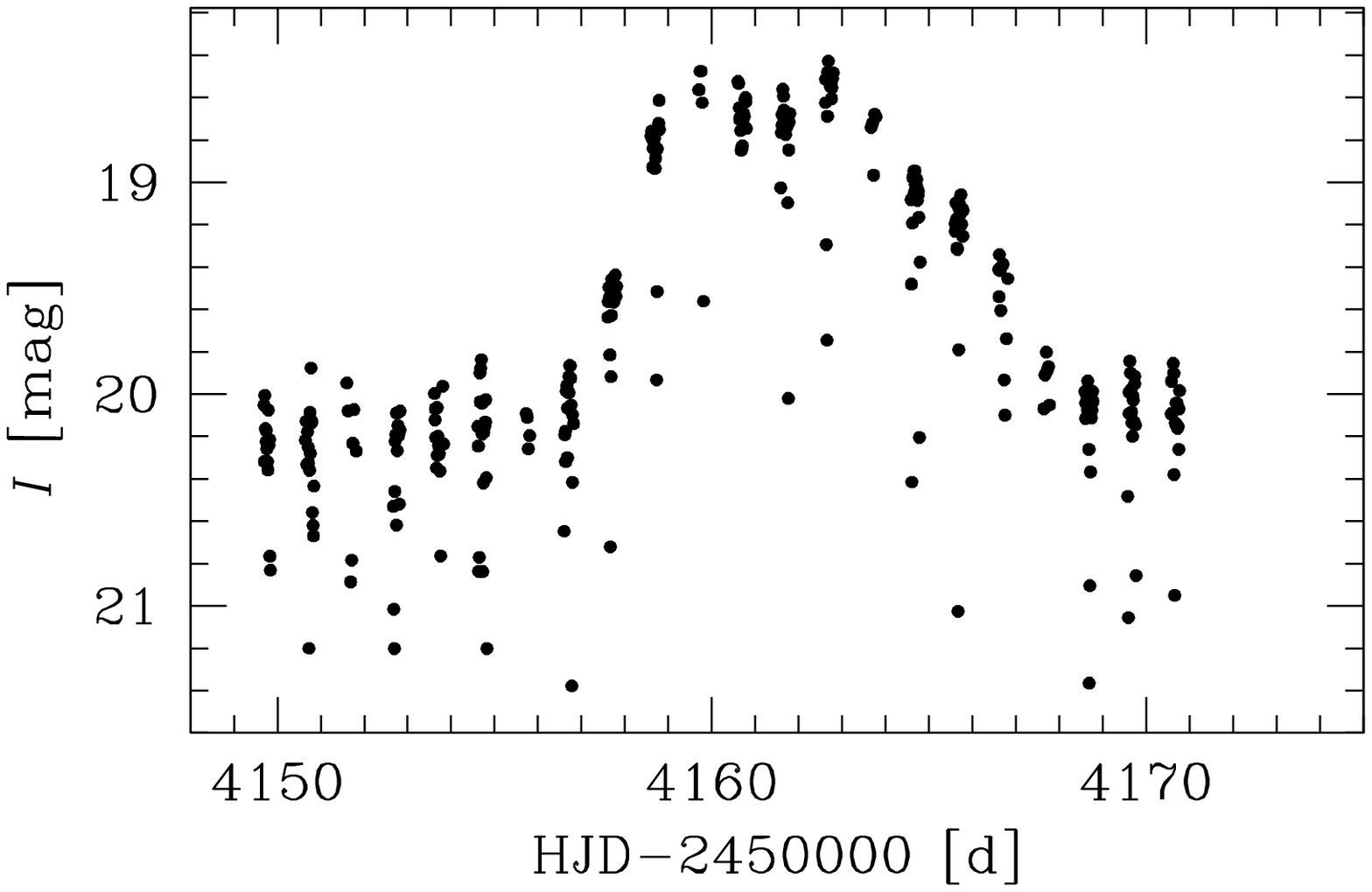}}
\FigCap{Close-up on the outburst with an embedded precursor
in OGLE-GD-DN-002.}
\end{figure}


\section{SU UMa- and WZ Sge type Stars}

In Table~3, we list basic photometric properties of stars identified
as SU~UMa- and WZ~Sge type variables: the mean brightness
in quiescence $I_{\rm qui}$, the mean maximum brightness in normal
outburst $I_{\rm n}$ and in superoutburst $I_{\rm s}$, the number
of observed normal outbursts $n_{\rm n}$ and superoutbursts $n_{\rm s}$,
the duration of normal outbursts $D_{\rm n}$ and superoutbursts $D_{\rm s}$,
the recurrence time between the normal outbursts $T_{\rm n}$,
the supercycle length $T_{\rm s}$, and the superhump period $P_{\rm sh}$,
if measured. Light curves of SU~UMa type variables, except for the WZ~Sge
type stars OGLE-GD-DN-001 and OGLE-GD-DN-014, are shown in Figs.~9--10.
Below we describe selected objects in detail.

\begin{table}
\centering
\caption{\small Photometric data on the SU~UMa- and WZ~Sge type variables}
\medskip
{\footnotesize
\begin{tabular}{ccccrrcrrrc}
\hline
Name & $I_{\rm qui}$ & $I_{\rm n}$ & $I_{\rm s}$ & $n_{\rm n}$ & $n_{\rm s}$ & $D_{\rm n}$ & $D_{\rm s}$ & $T_{\rm n}$ & $T_{\rm s}$ & $P_{\rm sh}$\\
 & [mag] & [mag] & [mag] & & & [d] & [d] & [d] & [d] & [d] \\
\hline
OGLE-GD-DN-001 & 21.63 & 19.24 &   17.80  &  4 &  1 & 2.0 &   18.0  &   -  &   -  & 0.06072(2) \\
OGLE-GD-DN-003 & 20.19 & 17.88 &   17.50  & 16 &  6 & 6.5 &   11.6  & 11.7 & 55.0 & - \\
OGLE-GD-DN-004 & 21.00 & 18.72 &   18.38  & 25 & 11 & 4.5 &    7.1  &  6.9 & 24.5 & - \\
OGLE-GD-DN-007 & 18.01 & 17.33 &   17.17  &  1 &  2 & 3.0 &   12.0  &   -  &   -  & 0.08081(4) \\
OGLE-GD-DN-008 & 22.05 & 18.16 &   17.82  &  2 &  1 & 4.5 &   18.0  &   -  &   -  & 0.08404(3) \\
OGLE-GD-DN-009 & 19.73 & 17.36 &   16.96  & 21 &  4 & 6.0 &   11.0  & 14.5 & 86.8 & 0.1310(3) \\
OGLE-GD-DN-014 & 22.65 & 19.68 & $>18.44$ &  2 &  1 & 3.5 & $>8.0$  &   -  &   -  & 0.08931(5) \\
OGLE-GD-DN-015 & 20.65 & 18.74 &   18.27  &  3 &  1 & 5.5 &   11.0  &   -  &   -  & - \\
OGLE-GD-DN-016 & 20.54 &   -   &   18.21  &  0 &  1 &  -  &   18.0  &   -  &   -  & - \\
OGLE-GD-DN-017 & 21.57 & 18.61 & $>18.47$ &  1 &  1 & 4.0 & $>9.0$  &   -  &   -  & - \\
OGLE-GD-DN-023 & 22.17 & 20.89 &   19.90  &  2 &  3 &  -  &   13.0  &   -  &   -  & - \\
OGLE-GD-DN-025 & 18.39 &   -   &   17.28  &  0 &  1 &  -  &   18.0  &   -  &   -  & - \\
OGLE-GD-DN-031 & 20.34 & 18.99 &   18.34  &  3 &  2 & 3.0 &   12.0  &   -  &   -  & 0.06324(1) \\
OGLE-GD-DN-033 & 22.08 & 19.69 &   19.32  &  6 &  3 & 4.6 &   10.0  & 18.0 &   -  & - \\
OGLE-GD-DN-034 & 20.98 & 18.50 &   17.43  &  5 &  2 & 5.0 & $>14.0$ &   -  &   -  & - \\
OGLE-GD-DN-036 & 21.91 & 20.08 &   19.69  & 13 &  8 & 4.4 &   10.3  & 11.4 & 55.3 & - \\
OGLE-GD-DN-037 & 16.19 &   -   &   15.88  &  0 &  3 &  -  &   14.8  &   -  &   -  & - \\
OGLE-GD-DN-038 & 20.76 &   -   &   19.19  &  0 &  1 &  -  &   14.0  &   -  &   -  & - \\
OGLE-GD-DN-039 & 19.76 & 18.63 &   18.20  & 35 &  5 & 2.7 &   13.0  &  4.5 & 80.0 & 0.08349(7) \\
\hline
\end{tabular}}
\end{table}

\begin{figure}
\centerline{\includegraphics[angle=0,width=100mm]{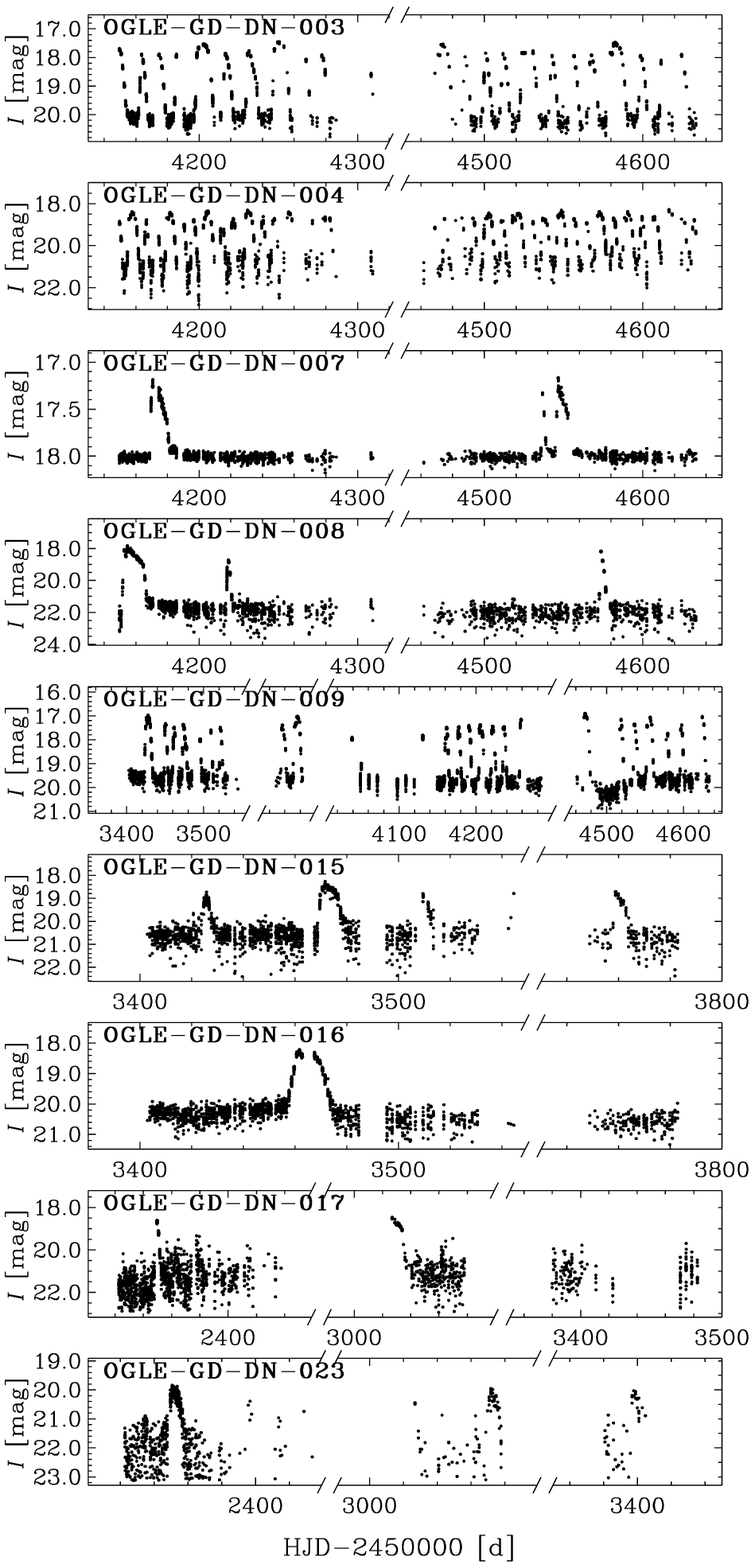}}
\FigCap{Light curves of the SU UMa type stars detected in the OGLE-III
disk fields (part 1 of 2). Small ticks on the time axis are every 20~d.}
\end{figure}

\begin{figure}
\centerline{\includegraphics[angle=0,width=100mm]{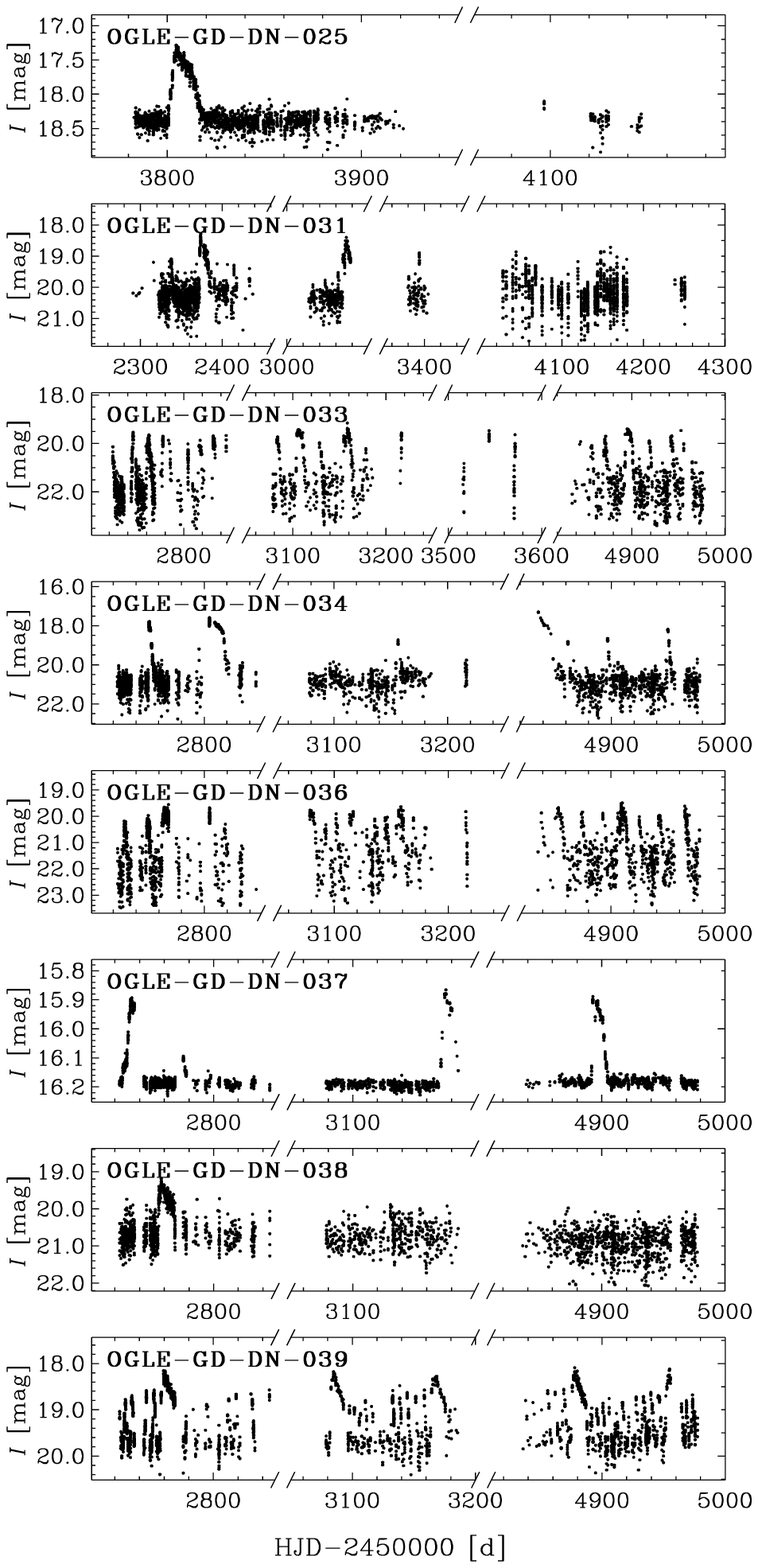}}
\FigCap{Light curves of the SU UMa type stars detected in the OGLE-III
disk fields (part 2 of 2). Small ticks on the time axis are every 20~d.}
\end{figure}

\subsection{OGLE-GD-DN-003}

A typical supercycle in this star lasts 55~d and contains
three normal outbursts. One of the supercycles, the one around
HJD=2454200, is shorter: 52~d with two normal outbursts.
During that supercycle normal outbursts lasted longer
($15.8\pm0.1$~d) in comparison to outbursts during the other
supercycles ($11.7\pm0.2$~d). Distinct precursor outbursts
were observed prior to superoutbursts. In quiescence, a periodic
signal with 0.15162(1)~d was found.

\subsection{OGLE-GD-DN-004}

This star has the shortest measured supercycle length in our
sample of SU~UMa type DNe. We found that the supercycle increased
from $24.5\pm0.1$ d in 2007 to $26.4\pm0.1$ d in 2008.
This corresponds to $\dot{P}_{\rm sc}=(+5.4\pm0.6)\cdot 10^{-3}$.
During the 2007 season the supercycles were very similar
to each other, while in 2008 the outbursts seemed to be
more erratic. We note the presence of precursors
before some superoutbursts.

\subsection{OGLE-GD-DN-007}

In this star only two superoutbursts were observed with distinct
superhumps with the period of 0.08081(4)~d. Possible lengths
of the supercycle are the following: 94~d, 125~d, 188~d or 376~d.

\subsection{OGLE-GD-DN-008}

In the light curve of this SU~UMa type variable only one
superoutburst was noted. The lower limit for the supercycle
length is about 160~d. Superhumps, already observed during
the precursor outburst, have the period of 0.08404(3)~d.
The superhump evolution is shown in Fig.~11.

\begin{figure}
\centerline{\includegraphics[angle=0,width=120mm]{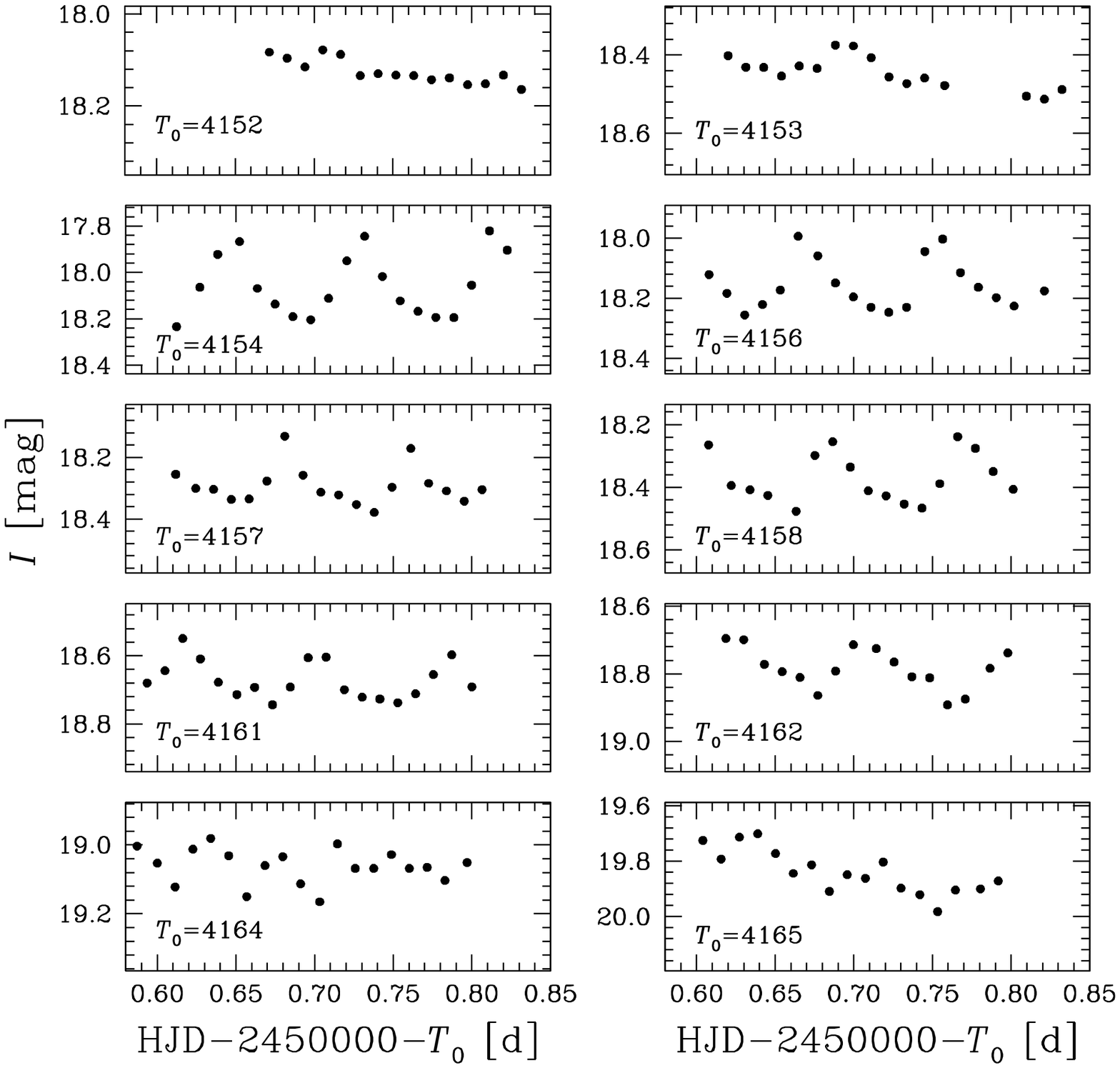}}
\FigCap{Evolution of superhumps in the SU~UMa type dwarf nova
OGLE-GD-DN-008 during the superoutburst in 2007.}
\end{figure}

\subsection{OGLE-GD-DN-009}

This star is located in two overlapping OGLE-III fields, CAR109 and CAR115,
and was monitored for a record number of 379 nights over seasons 2005--2008.
Four long and bright superoutbursts with distinct precursors and
superhumps were observed. The estimated supercycle period amounts
to 86.8~d, while the recurrence time of the normal outbursts increased
systematically from 12~d to 19~d. The superhump period is $0.1310(3)$~d.

\subsection{OGLE-GD-DN-023}

This star is one of the faintest stars in our sample.
Three bright superoutbursts and two normal outbursts
were observed. The possible supercycle length is either
65.5~d or 80~d. During the first superoutburst we found
low-amplitude variations (probably superhumps)
with the period of either 0.0705(5)~d or 0.0755(5)~d.

\subsection{OGLE-GD-DN-034}

Only two superoutbursts were observed for this object. During the
first one, superhumps with the period of $0.06324(1)$~d were found.
The shortest plausible supercycle length for this star
is 70~d, however other values are also possible.

\subsection{OGLE-GD-DN-036}

The supercycle length in this object increased from $43.7\pm0.3$~d
in 2003 to $55.3\pm0.7$~d in 2004. This corresponds to
$\dot{P_{\rm sc}} = (+5.9\pm0.4)\cdot 10^{-3}$. The recurrence time
between the normal outbursts varies significantly, from 7~d to 18~d
with a mean value of 10--11~d.

\subsection{OGLE-GD-DN-039}

The light curve of this stars shows five bright ($I_{max}\approx18.2$~mag)
and long ($\approx13$~d) superoutbursts and 10 normal outbursts
in between ($I_{\rm max}\approx18.6$~mag, $2.7\pm0.2$~d).
Although no precursor outbursts were observed (unluckily, due to
bad weather conditions the beginning of two superoutbursts were
missed), the presence of superhumps indicates that this object is
a {\it bona fide} SU~UMa type variable. The superhump period amounts
to $P_{\rm sh}=0.08349(7)$~d $=2.004(2)$~h. The average supercycle length
is 80.0(1)~d, however, it decreased from 81.5~d to 78.4~d with a constant
rate of $\dot{P}_{\rm sc}=-1.36(8) \cdot 10^{-3}$. This is in contradiction
to what has been recently found in active SU~UMa type stars
(Otulakowska-Hypka \etal 2013b). The object OGLE-GD-DN-039 seems
to be one of the missing systems between the typical SU~UMa type
variables and ER~UMa type stars.

\subsection{OGLE-GD-DN-001 and OGLE-GD-DN-014}

These two WZ~Sge type stars have similar light curves, shown in Fig.~12.
In both cases we observed bright and long superoutburst with distinct
superhumps followed by a series of 2--4 short ``echo'' outbursts.
The $I$-band amplitude of the superoutbursts are 2.9~mag and 4.2~mag
in OGLE-GD-DN-001 and OGLE-GD-DN-014, respectively, while the ``echo''
outbursts are $\approx1.4$~mag fainter, in both stars.
Unfortunately, the beginnings of the superoutbursts were not observed.
In OGLE-GD-DN-001 the points around HJD=2454170 may correspond to the
precursor outburst. We assessed the superhump period for each star.
For OGLE-GD-DN-001 it is equal to 0.06072(2)~d $=1.4572(3)$~h.
The second star, OGLE-GD-DN-014, has a slightly longer superhump
period, $P_{\rm sh}$, of 0.08931(5)~d $=2.1434(10)$~h.

These two stars are very similar in their behavior to EG~Cnc which
was observed in superoutburst in 1996 (Patterson \etal 1998). This
suggests that OGLE-GD-DN-001 and OGLE-GD-DN-014 are WZ~Sge type stars
with a very small mass-transfer rate and extremely long supercycles.
Two similar systems, namely MBR124.21.11N and MBR120.21.138N, were recently
discovered in the OGLE Magellanic Bridge region (Koz{\l}owski \etal 2013).
Future OGLE observations should allow estimation of the supercycle
lengths in all these objects.

\begin{figure}
\centerline{\includegraphics[angle=0,width=130mm]{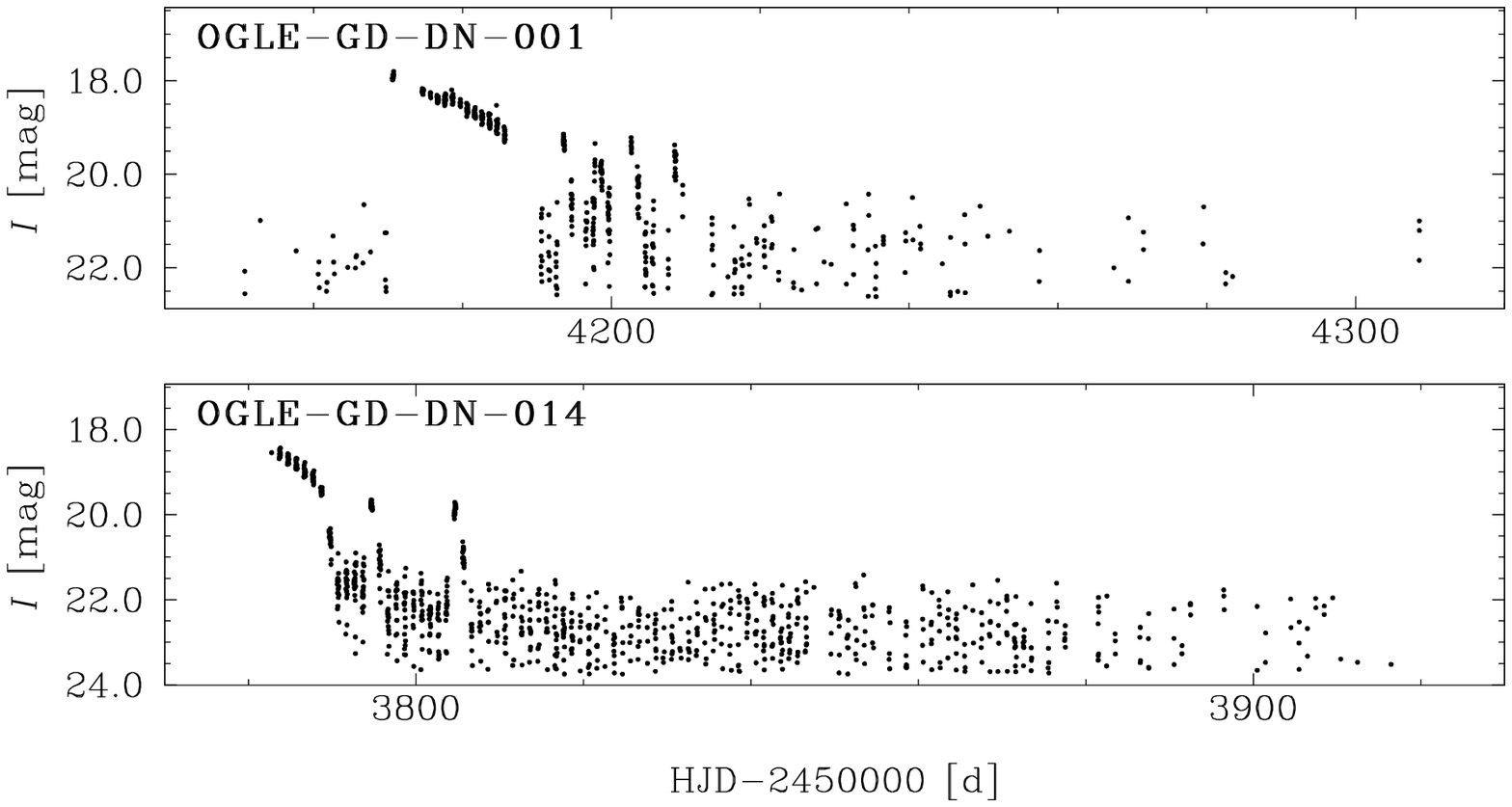}}
\FigCap{Light curves of two WZ~Sge type stars. Note the presence
of ``echo'' outbursts after the superoutbursts.}
\end{figure}


\section{Z Cam type Stars}

Basic photometric properties of four Z~Cam type stars in the OGLE-III
disk area are listed in Table~4. We measured the following values:
the mean brightness in standstill $I_{\rm stand}$, the maximum
$I_{\rm max}$ and minimum $I_{\rm min}$ brightness in outburst,
the duration of outbursts $D_{\rm out}$, and the recurrence time
between outbursts $T_{\rm out}$. Light curves of the stars are plotted
in Fig.~13. The detected Z~Cam type stars spent on average 25\% of the
time in standstills. Below we give detail comments on each object.

\begin{table}
\centering
\caption{\small Photometric data on the newly discovered Z~Cam type stars}
\medskip
\begin{tabular}{ccccrr}
\hline
Name & $I_{\rm stand}$ & $I_{\rm max}$ & $I_{\rm min}$ & $D_{\rm out}$ & $T_{\rm out}$\\
 & [mag] & [mag] & [mag] & [d] & [d]\\
\hline
OGLE-GD-DN-012 & 17.71 & 17.13 & 18.28 & 6-8 & 9 \\
OGLE-GD-DN-027 & 19.15 & 18.35 & 19.71 & 8-15 & 20\\
OGLE-GD-DN-028 & 19.70 & 18.37 & 20.15 & 6-16 & 26\\
OGLE-GD-DN-035 & 18.73 & 17.62 & 20.63 & 7-16 & 17-20\\
\hline
\end{tabular}
\end{table}

\begin{figure}
\centerline{\includegraphics[angle=0,width=130mm]{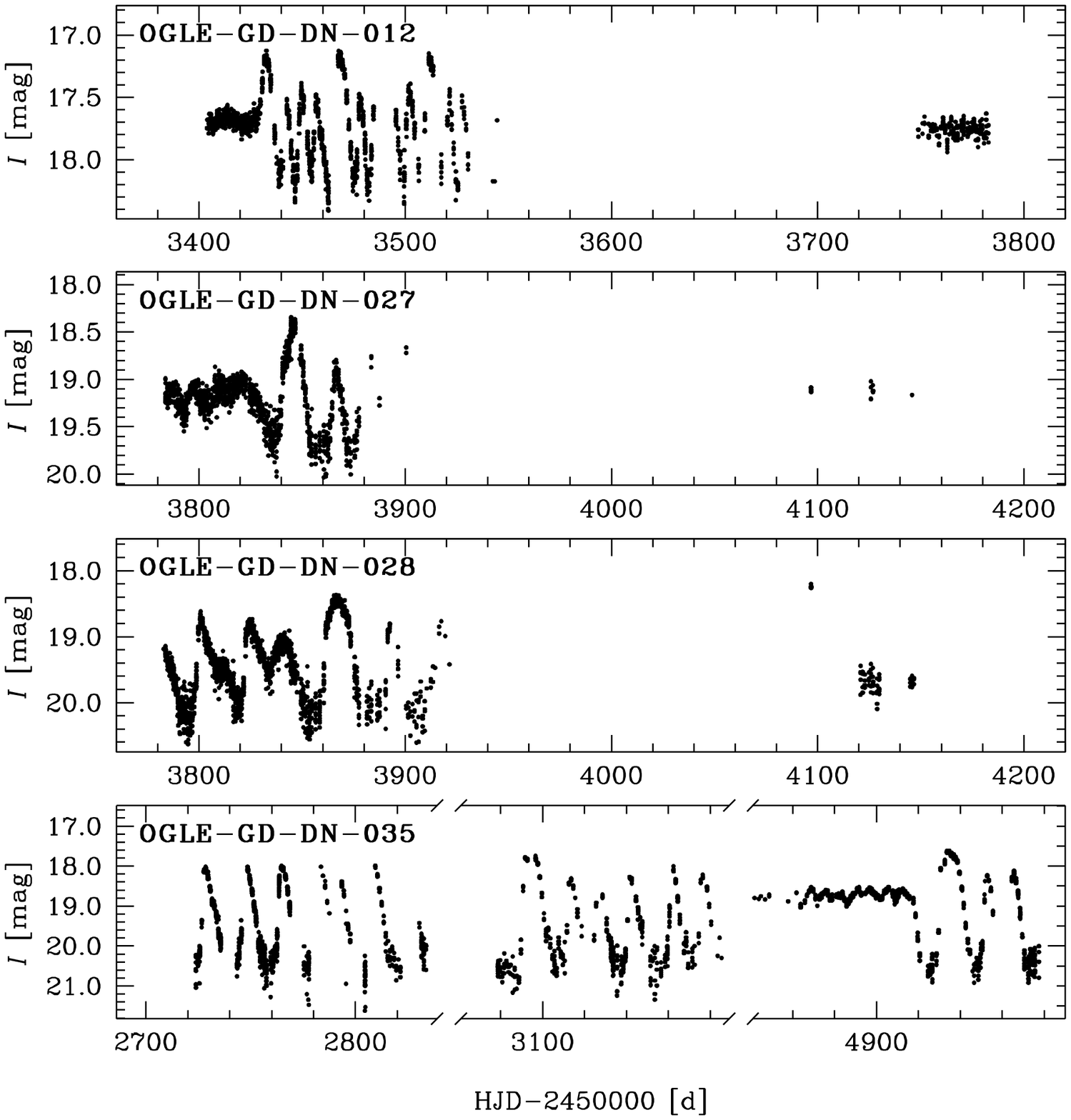}}
\FigCap{Light curves of Z Cam type stars.}
\end{figure}

\subsection{OGLE-GD-DN-012}

This star was observed for two seasons (2005--2006). In its light curve,
one can see fainter ($I_{\rm max}\approx17.5$~mag, amp$_I\approx0.6$~mag)
and shorter (5--6~d) regular outbursts interlaced every 3--4 events
with more prominent ($I_{\rm max}\approx17.1$~mag, amp$_I\approx1.1$~mag)
and longer (10--11~d) outbursts. In minimum, the $I$-band brightness
is 18.3~mag and is similar after both types of outbursts. It is worth
to note that, in contrast to the majority of known Z~Cam type stars,
the brightness in OGLE-GD-DN-012 increased after the first standstill.
Unfortunately, the end of the second standstill was not observed.

\subsection{OGLE-GD-DN-027}

We observed one clear standstill for this star. The first outburst
after the standstill was brighter and longer ($I_{\rm max}\approx18.35$~mag,
15~d) than the following three outbursts ($I_{\rm max}\approx18.80$~mag,
8~d). During the standstill the brightness varied with an amplitude of
$\approx0.5$~mag and a time scale of $\approx12$~d.

\subsection{OGLE-GD-DN-028}

In this object two very short standstills were observed,
around HJD=2453810 and HJD=2454120. The $I$-band amplitudes of
1.2--2.0~mag and irregular shape of the outbursts are typical
for Z~Cam type stars.

\subsection{OGLE-GD-DN-035}

This is the best observed Z~Cam type star in our sample.
Its light curve covers 279 nights over years 2003--2009.
It includes 15 outbursts and one standstill lasting
at least 60~d. In the first two seasons, the star
exhibited outbursts recurring every 17--20~d.
The first outburst after the standstill was brighter
and longer ($I_{\rm max}\approx17.6$~mag, 16~d) than the
following ones ($I_{\rm max}\approx18.1$~mag, 7~d).
OGLE-GD-DN-035 was observed in the standstill phase for about 22\%
of the time. In this state the brightness variations
with an amplitude of $\approx0.4$~mag and a time scale
of $\approx7.5$~d were seen.


\section{Summary and Conclusions}

In this paper we report the identification of forty new dwarf novae
in twenty-one OGLE-III Galactic disk fields. We have increased the total
number of all known DNe by 7 per cent. Seventeen variables
are of the U~Gem type, nineteen of the SU~UMa type, and four
of the Z~Cam type. We analyzed the light curves and estimated basic
parameters of the DNe such as the duration and recurrence period of the
outbursts. Prominent eclipses seen in OGLE-GD-DN-002 allowed us to find
precise orbital period in this system. We note that in one of
our Z~Cam type objects, OGLE-GD-DN-012, in contrast to the
large majority of variables of this type, we observed an increase
of brightness after the standstill phase.

In the case of five DNe classified as SU~UMa type variables
we were able to assess supercycle lengths. The obtained lengths
are in the range 20--90~d, generally between objects
classified as typical SU~UMa type and ER~UMa type variables.
In Fig.~14, we present the distribution of the supercycle lengths
for 73 known SU~UMa type stars, including the five new objects.
The number of detected short supercycle period DNe is large
enough to discard the ER~UMa type as a separate class of variables
and to classify them as SU~UMa type variables. We note that in object
OGLE-GD-DN-039 we found a decrease in the supercycle period --- not
observed in other active SU~UMa type systems.

Currently, there is no known DN with a supercycle period between
1100~d and 3300~d. Dozens of systems with a single observed
WZ~Sge-like superoutburst require follow-up monitoring campaigns
with the aim to find their $P_{\rm sc}$. In this paper we add two stars,
OGLE-GD-DN-001 and OGLE-GD-DN-014, to the list of such systems.
Analysis of time-series photometry from exisiting (\eg 21-year-long
bulge observations by OGLE) and future long-term surveys (\eg
Large Synoptic Survey Telescope, LSST Science Collaboration 2009)
should allow statistical assessment of the existence of objects
with $P_{\rm sc}$ in this range.

\begin{figure}
\centerline{\includegraphics[angle=0,width=130mm]{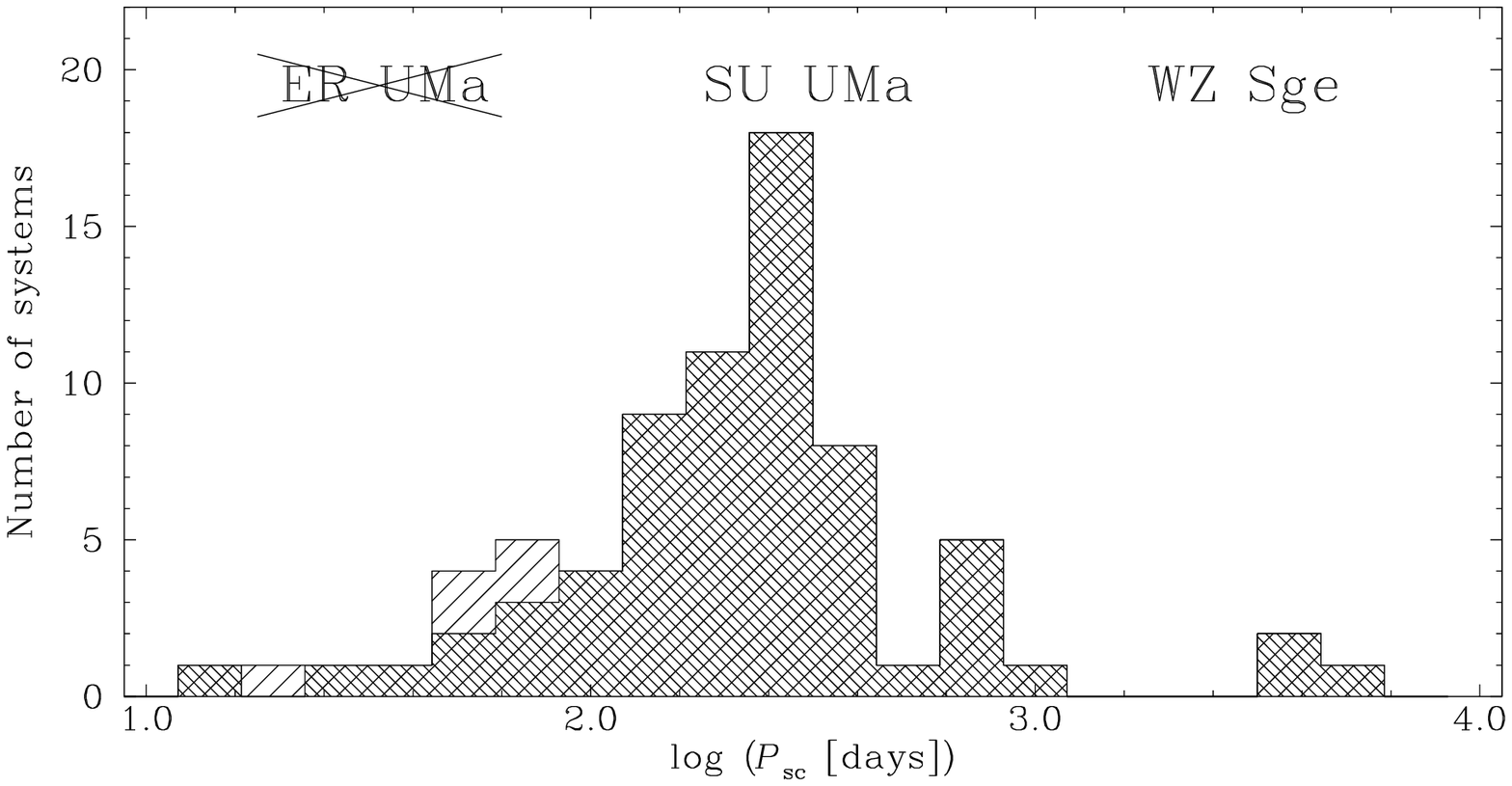}}
\FigCap{Distribution of the supercycle lengths for known SU~UMa type
stars. Five OGLE stars have, more or less, supercycle lengths between
the short period ER~UMa type stars and the typical SU~UMa type DNe.
Thus, we propose to discard the former type.}
\end{figure}


\Acknow{
This work has been supported by the Polish National Science Centre
grant No. DEC-2011/03/B/ST9/02573 and the Polish Ministry of Sciences
and Higher Education grant No. IP2012 005672 under Iuventus Plus
programme.

The OGLE project has received funding from the European
Research Council under the European Community$'$s Seventh Framework
Programme (FP7/2007-2013)/ERC grant agreement No. 246678 to A.U.}


\end{document}